\documentclass[preprint, prd, nofootinbib, aps, showpacs, showkeys, preprintnumbers]{revtex4-1}
\pdfoutput=1
\usepackage{amsmath,amsfonts,amssymb,amsthm,amstext,amscd,eucal}

\usepackage[all]{xy}
\usepackage{epsfig}
\usepackage{amsmath}
\usepackage{amssymb}
\usepackage{graphicx}

\long\def\symbolfootnote[#1]#2{\begingroup
\def\thefootnote{\fnsymbol{footnote}}
\footnote[#1]{#2}\endgroup}

\makeatletter \@addtoreset{equation}{section}

\numberwithin{equation}{section}

\newcommand{\ie}{{\em i.e.~}}

\newcommand{\etal}{\emph{et\,al.~}}

\newcommand{\be}{\begin{equation}}  \newcommand{\ee}{\end{equation}}
\newcommand{\bea}{\begin{eqnarray}} \newcommand{\eea}{\end{eqnarray}}
\newcommand{\nn}{\nonumber}

\newcommand{\bse}{\begin{subequations}}
\newcommand{\ese}{\end{subequations}}
\newcommand{\bi}{\begin{itemize}}
\newcommand{\ei}{\end{itemize}}
\newcommand{\lp}{\left(}
\newcommand{\rp}{\right)}

\newcommand{\mpl}{M_{\rm Pl}}

\newcommand\dsigma{\dot\sigma}
\newcommand\ddsigma{\ddot\sigma}
\newcommand\dalpha{\dot\alpha}
\newcommand\ddalpha{\ddot\alpha}
\newcommand\dbeta{\dot\beta}
\newcommand\ddbeta{\ddot\beta}
\newcommand\dpsi{\dot\psi}
\newcommand\ddpsi{\ddot\psi}
\newcommand\dchi{\dot\chi}
\newcommand\ddchi{\ddot\chi}

\begin{document}
\preprint{IPM/P-2013/042}
\vspace{2cm}
\begin{center}
\large{\bf Chromo-Natural Model in Anisotropic Background}
\end{center}
\begin{center}
\large{Azadeh Maleknejad$^{*}$} and \large{Encieh Erfani$^{\dag}$}
\end{center}
\begin{center}
\textit{School of Physics, Institute for Research in Fundamental Sciences (IPM),\\ P. Code. 19538-33511, Tehran, Iran}
\end{center}

\symbolfootnote[0]{$^{*}$azade@ipm.ir}
\symbolfootnote[0]{$^{\dag}$eerfani@ipm.ir}

\begin{abstract}
In this work we study the chromo-natural inflation model in the anisotropic setup. Initiating inflation from Bianchi type-I cosmology, we analyze the system thoroughly during the slow-roll inflation, from both analytical and numerical points of view. We show that the isotropic FRW inflation is an attractor of the system. In other words, anisotropies are damped within few $e$--folds and the chromo-natural model respects the cosmic no-hair conjecture. Furthermore, we demonstrate that in the slow-roll limit, the anisotropies in both chromo-natural and gauge-flation models share the same dynamics. 
\end{abstract}

\maketitle

\section{Introduction}

The early epoch of almost exponential expansion, known as inflation, was proposed to resolve some of standard Big Bang cosmology puzzles such as the horizon and the flatness problems \cite{Guth81}. Besides these remarkable successes, inflation is also able to describe the cosmological data and CMB temperature anisotropies \cite{Planck}. Moreover, the inflationary paradigm has the powerful theoretical and model building feature of almost relaxing the dependence of late time physics on the pre-inflation initial conditions \cite{Inflation-texts}-\cite{Bassett}. One particular set of such conditions is anisotropic but homogeneous metric; the Bianchi cosmology models. In this work, we are interested in this family of pre-inflationary initial conditions and in particular Bianchi type I.

For inflationary models consist of (only) scalar fields, the anisotropic part of the metric is not sourced by the (scalar) matter field. Hence, anisotropies are (exponentially) damped by the inflationary expansion of the background. However, this picture may change in the presence of vector (gauge) fields, where the gauge field can source the anisotropic part of the metric. More precisely, \textit{inflationary extended cosmic no-hair theorem} \cite{Maleknejad:2012as} states that for general inflationary systems of all Bianchi types, anisotropies can grow during inflation, in contrast to the cosmic no-hair conjecture (and also Wald's cosmic no-hair \cite{Wald:1983ky}). Nevertheless, inflation puts an upper bound on the enlargement of the anisotropies. Assuming a (slow-roll) quasi-de Sitter expansion, the enlargement of anisotropies is at most of order of the slow-roll parameter $\epsilon$ \cite{Maleknejad:2012as}.

Introducing a general class of non-Abelian gauge field inflationary models minimally coupled to Einstein gravity, gauge-flation opens a new venue for building inflationary models closer to particle physics models. In \cite{Maleknejad:2011jw, Maleknejad:2011sq}, it was shown that non-Abelian gauge field theory can provide a setup for constructing isotropic and homogeneous inflationary background. For an extensive review on this topic see \cite{Maleknejad:2012fw}. Another interesting inflationary scenario involving non-Abelian gauge fields is the chromo-natural model \cite{Adshead:2012kp}. In this model, inflation is driven by an axion field, while the non-Abelian gauge field plays the important role in flattening the axion potential, without requiring (unnatural) super-Planckian axion decay constant \cite{Adshead:2012kp}-\cite{Dimastrogiovanni:2012st}. In other words, the axion-gauge field interaction makes the model technically natural. It is worth to mention that both gauge-flation and chromo-natural models have been disfavored by recent CMB observational data \cite{Planck, Namba:2013kia, Adshead:2013nka}.

Due to the vector nature of the gauge fields, inflationary models involving gauge fields may or may not respect the cosmic no-hair conjecture \cite{Maleknejad:2012as, Soda:2012zm}. Since the gauge field sources anisotropies in (homogeneous but anisotropic) Bianchi background, it is not clear that the isotropic setup is a stable solution in these models. Hence, it is necessary and important to study the stability of the isotropic background against the initial anisotropies. This matter has been studied in \cite{Maleknejad:2011jr} for the gauge-flation model. There, it was shown that although the gauge field is turned on in the background, the isotropic configuration is an attractor solution of the model and the initial anisotropies are damping within few $e$--folds. Considering Bianchi type I, we tackle the same issue in the chromo-natural model and we study the classical stability of this model against the initial anisotropies during the slow-roll inflation.

The rest of this paper is organized as follows. In Sec.~\ref{II}, we study the chromo-natural model in Bianchi type I cosmology. In Sec.~\ref{III}, we analyze the system during the slow-roll inflation, both from analytical and numerical points of view. Finally, we conclude in Sec.~\ref{V}.

Throughout this note $\mpl=(8\pi G)^{-1/2}=1$.

\section{Setup}\label{II}

Consisting of an axion field $\chi$ and a non-Abelian $SU(2)$ gauge field $A_{~\mu}^a$, the chromo-natural model \cite{Adshead:2012kp} is an inflationary model with the following action
\bea\label{action1}
S=\int d^4x\,\sqrt{-g}\,\left[-\frac{R}{2}-\frac{1}{4}F_{~\mu\nu}^a\,F^{~\mu\nu}_a-\frac{1}{2}(\partial_{\mu}\chi)^2-\mu^4\lp 1+\cos(\frac{\chi}{f})\rp+\frac{\lambda}{8}\dfrac{\chi}{f}F^a_{~\mu\nu}\tilde{F}_a^{~\mu\nu}\right]\,,
\eea
where $\tilde F_a^{~\mu\nu}=\epsilon^{\mu\nu\lambda\sigma} F_{~\lambda\sigma}^a$ and the gauge field strength tensor $F_{~\mu\nu}^a$, is given by
\be
F_{~\mu\nu}^a=\partial_\mu A_{~\nu}^a-\partial_\nu A_{~\mu}^a-g\,\epsilon^a_{~bc}A_{~\mu}^b A_{~\nu}^c\,.
\ee
Here $a,\,b,\,c=1,\,2,\,3$ are used for the indices of algebra while $\mu,\,\nu=0,\,1,\,2,\,3$ represent the space-time indices. The above action has 4 parameters, dimensionless parameters $\lambda$ and $g$ and dimensionful parameters $\mu$ and $f$, where $f$ is the axion decay constant.

The homogeneous and \textit{isotropic} FRW cosmology of the chromo-natural model has been widely studied in \cite{Adshead:2012kp}-\cite{Maleknejad:2014wsa}. In this model, the axion field is the inflaton, while the non-Abelian gauge field is a secondary field which is coupled to the axion through its Chern-Simons interaction. In the absence of the non-Abelian gauge field, this model reduces to the natural inflation \cite{Adams:1992bn, Freese:2004un}. In this model, the slow-roll inflation is obtained for super-Planckian $f$ parameter, which is not a natural scale within particle physics models. However, in the chromo-natural model, the non-Abelian gauge field slows down the inflaton's evolution and leads to slow-roll inflation even with the natural values of $f$ ($f\ll \mpl$).

In the FRW background, the gauge field has the following homogeneous and isotropic solution
\bea\label{ansatz-iso}
A_{~\mu}^{a}|_{_{\rm FRW}}=\left\{
\begin{array}{ll}
0                                                          &\quad \mu=0\,,\\
\psi(t) {\sf e}_{~i}^a|_{_{\rm FRW}}=a(t)\psi(t)\delta^a_i &\quad \mu=i\,,
\end{array}
\right.
\eea
where ${\sf e}_{~i}^a|_{_{\rm FRW}}$ are the spatial triads of the FRW metric, $a(t)$ is the scale factor and $\psi$ is a scalar field. Using the axion field equation and imposing the slow-roll condition, $\psi$ is determined in terms of $\chi$ as following
\be\label{psi-sl}
\psi|_{_{\rm FRW}}\simeq\lp\dfrac{\mu^4\sin(\chi/f)}{3\lambda g H}\rp^{1/3}\,,
\ee
where $H$ is the Hubble parameter. Note that in the homogeneous and isotropic FRW background, $H$, $\chi$ and $\psi$ are evolving slowly during the slow-roll inflation. For a thorough review on this issue see \cite{Maleknejad:2012fw}.

Since the chromo-natural model includes a non-Abelian gauge field in the background, a question which may arise naturally is the classical stability of the isotropic inflationary trajectory against initial anisotropies. Since the gauge field sources anisotropies in (homogeneous but anisotropic) Bianchi background, it is not clear if the isotropic setup is a stable solution. Here, considering the chromo-natural model in the homogeneous but \textit{anisotropic} Bianchi type I background, we study the behavior of anisotropies during inflation.

The space-time metric in the homogeneous but \textit{anisotropic} (axially symmetric) Bianchi type I metric is given by

\be\label{metric}
ds^2=-dt^2+e^{2\alpha(t)}\left(e^{-4\sigma(t)}dx^2+e^{2\sigma(t)}(dy^2+dz^2)\right)\,,
\ee
where $e^{\alpha(t)}$ is the isotropic scale factor and the anisotropy in the metric is represented by (the time derivative of) $\sigma(t)$.
Choosing the temporal gauge, the consistent truncation for the gauge field reads as
\bea\label{ansatz}
A_{~\mu}^{a}=\left\{
\begin{array}{ll}
0                      &\mu=0\,,\\
\psi_i {\sf e}_{~i}^a  &\mu=i\,,
\end{array}
\right.
\eea
where the spatial triads have the following explicit forms
\be
{\sf e}_{~1}^a=e^{\alpha-2\sigma}\delta^a_{~1}\,, \quad  {\sf e}_{~2}^a=e^{\alpha+\sigma}\delta^a_{~2} \, \quad \textmd{and}\quad {\sf e}_{~3}^a=e^{\alpha+\sigma}\delta^a_{~3}\,.\nn
\ee
Note that, due to the axial symmetry of the metric in $y-z$ plane, we have $\psi_2=\psi_3$.
As a result, the explicit form of our ansatz is given by
\be
A_{~i}^a=\text{diag}\,(e^{\alpha-2\sigma}\psi_1,\,e^{\alpha+\sigma}\psi_2,\,e^{\alpha+\sigma}\psi_2)\,.\nn
\ee
For mathematical convenience, which will be clear soon, we introduce the following field redefinitions
\be
\psi_1(t)=\dfrac{\psi(t)}{\beta(t)^2}\,; \quad \psi_2(t)=\beta(t)\psi(t)\,,
\ee
where $\psi$ represents the \textit{isotropic} field and $\beta$ parametrizes \textit{anisotropy} in the gauge field.
Comparing Eq.~\eqref{ansatz-iso} and Eq.~\eqref{ansatz}, we realize that the isotopic solution is given by $\beta^2=1$. Thus the deviation of $\beta^2$ from one parametrizes the amount of anisotropy in the gauge field.

Determining the energy-momentum tensor and plugging the metric \eqref{metric} and the ansatz \eqref{ansatz} into it, we have a diagonal homogeneous tensor
$$T^{\mu}_{~\nu}={\rm diag}\,(-\rho,\,P-2\tilde P,\,P+\tilde P,\,P+\tilde P)\,,$$
in which $\rho$ is the energy density, while $P$ and $\tilde P$ are the isotropic and anisotropic pressures, respectively. By decomposing the energy density as $\rho=\rho_\chi+\rho_{_{\rm YM}}$, where $\rho_\chi$ and $\rho_{_{\rm YM}}$ are contributions of axion field and Yang-Mills term respectively\footnote{Since the Chern-Simons interaction is a topological term, it does not provide any contribution to the energy-momentum tensor.}, we have
\bea\label{k-rho}
\rho_\chi&=&\frac{1}{2}\dot\chi^2+\mu^4\lp 1+\cos(\frac{\chi}{f})\rp\,,\\
\label{ym-rho}
\rho_{_{\rm YM}}&=&\frac{1}{2\beta^4}\lp\dot\psi+(\dot\alpha-2\dot\sigma-2\dfrac{\dot\beta}{\beta})\psi\rp^2+\beta^2\lp\dot\psi+(\dot\alpha+\dot\sigma-\dfrac{\dot\beta}{\beta})\psi\rp^2+g^2\dfrac{(2+\beta^6)}{2\beta^2}\psi^4\,.
\eea
Isotropic and anisotropic pressures are given by
\bea\label{pressure}
P&=&\dot\chi^2-\rho_\chi+\frac{1}{3}\rho_{_{\rm YM}}\,,\\
\label{P2}
\tilde P&=&\dfrac{(1-\beta^6)}{3}\lp\dfrac{1}{\beta^4}\big(\dot\psi+(\dot\alpha-2(\dot\sigma+\frac{\dot\beta}{\beta}))\psi\big)^2-\frac{1}{\beta^2}g^2\psi^4\rp-\beta^2(\dfrac{\dot\beta}{\beta}+\dot\sigma)\psi\nn\\
&\times&\bigg(2\dot\psi+\big(2\dot\alpha-(\dot\sigma+\frac{\dot\beta}{\beta})\big)\psi\bigg)\,.
\eea
Interestingly, the anisotropic pressure $\tilde P$, is only originated from the Yang-Mills term and the other terms do not contribute. Note that, for the isotropic case in which $\beta^2=1$ (where $\dot{\sigma}=0$ and $\dot\beta=0$), the anisotropic pressure vanishes.

From the Einstein equations, the inflation condition $(e^\alpha\ddot{)}>0$ leads to
\be
\frac{1}{6}(\rho+3P)+2\dot\sigma^2\leq0\,,
\ee
which implies that the inflation happens only if the axion potential $V(\chi)=\mu^4\bigg(1+\cos(\frac{\chi}{f})\bigg)$, is the dominant term in the energy density.

At this point, inserting the metric \eqref{metric} and the ansatz \eqref{ansatz} into the action \eqref{action1}, one obtains the explicit form of the action
\bea\label{action2}
S&=&\int d^4x\,e^{3\alpha}\bigg[-3\dalpha^2+3\dsigma^2+\frac{1}{2\beta^4}\bigg(\dpsi+(\dalpha-2\dsigma-2\frac{\dbeta}{\beta})\psi\bigg)^2+\beta^2\bigg(\dpsi+(\dalpha+\dsigma+\frac{\dbeta}{\beta})\psi\bigg)^2\nn\\
&-&\frac{g^2}{\beta^2}\psi^4-\dfrac{g^2}{2}\beta^4\psi^4+\frac{1}{2}\dchi^2-\mu^4\lp 1+\cos(\frac{\chi}{f})\rp-3\lambda g\psi^2(\dpsi+\dalpha\psi)\frac{\chi}{f}\bigg]\,.
\eea
Interestingly, the fields $\dot\sigma$ and $\beta$ only appear in the Yang-Mills term of the matter action.

Since the axion field $\chi$, couples only to $\psi$, its field equation is exactly the same one in the isotropic background
\bea\label{chieq}
\ddchi+3\dot\alpha\dchi-\frac{\mu^4}{f}\sin(\frac{\chi}{f})=-3g\frac{\lambda}{f}\psi^2(\dot\alpha\psi+\dpsi)\,.
\eea
As a result, the non-Abelian gauge field slows down the inflaton's evolution through its Chern-Simons interaction with the axion and leads to slow-roll inflation. Imposing the slow-roll conditions in the above equation, we obtain the slow-roll trajectory of $\psi$ in terms of $\chi$
\be\label{psisl}
\psi\simeq\lp\dfrac{\mu^4\sin(\chi/f)}{3\lambda g\,\dot\alpha}\rp^{1/3}\,.
\ee
The field equation of $\psi$ is given by

On the other hand, the field equation of $\psi$ is different in the anisotropic setup and is given by
\bea\label{psieq}
&~&\ddpsi\beta^2(2\beta^6+1)+\dpsi\bigg( 4\dbeta\beta(\beta^6-1)+3\dalpha\beta^2(2\beta^6+1)\bigg)+\psi\bigg(\ddalpha\beta^2(2\beta^6+1)+2\ddsigma\beta^2(\beta^6-1)+\nn\\
&~&2\ddbeta\beta(\beta^6-1)+6\dbeta^2+6\dalpha\dbeta\beta(\beta^6-1)+2\dalpha^2\beta^2(2\beta^6+1)+2\dalpha\dsigma\beta^2(\beta^6-1)-2\dsigma^2\beta^2(\beta^6+2)\bigg)\nn\\
&~&-3\dfrac{\lambda g}{f}\beta^6\dchi\psi^2+2g^2\beta^4(\beta^6+2)\psi^3=0\,,
\eea
which is coupled to $\beta$.

Then, from the action \eqref{action2}, one can determine $\dsigma$,
\be\label{sigmaeq}
\dsigma=\dot{\alpha}\psi^2\bigg(\dfrac{\frac{\dpsi}{\dot{\alpha}\psi}(1-\beta^6)-\frac{\dbeta}{\dot{\alpha}\beta}(2+\beta^6)+(1-\beta^6)}{3\beta^4+\psi^2(2+\beta^6)}\bigg)\,,
\ee
and the field equation of $\beta$
\bea\label{betaeq}
&~&\beta(\beta^6+2)\bigg(\ddbeta+3\dalpha\dbeta+2\dbeta\dfrac{\dpsi}{\psi}\bigg)\psi+\beta^2(\beta^6-1)\bigg( \ddpsi+3\dalpha\dpsi+(\ddalpha+2\dalpha^2)\psi+g^2\beta^2\psi^3\bigg)\nn\\
&~&+\beta^2(\beta^6+2)(\ddsigma+\dalpha\dsigma)\psi-6\dbeta^2\psi-\beta^2(\beta^6-4)\dsigma^2\psi=0\,.
\eea

Up to this point, we worked out the field equations and investigated the inflation condition in the anisotropic setup. 
In the next section we study the behavior of the anisotropic system during the slow-roll regime.

\section{Slow-roll regime}\label{III}

Slow-roll inflation is quantified in terms of slow-roll parameters, $\epsilon$ and $\eta$
\be\label{srp}
\epsilon= -\dfrac{\dot H}{H^2}\,, \quad  \eta=-\dfrac{\ddot H}{2H\dot H}\,,
\ee
where $H=\dot{\alpha}$ is the Hubble parameter and slow-roll conditions require $\epsilon,\,|\eta|\ll 1$.
Using the Einstein equations, the first slow-roll parameter $\epsilon$ is 
\be\label{epsilon-rho}
\epsilon=\frac{(\rho+P)/2+3\dot{\sigma^2}}{\rho/3+\dot{\sigma^2}}\,,
\ee
which demanding the slow-roll inflation, leads to following conditions
\be\label{sl-V}
V(\chi)\gg\dchi^2\,,\quad V(\chi)\gg\rho_{_{\rm YM}}\,, \quad\textmd{and}\quad V(\chi)\gg \dot{\sigma^2}\,.
\ee
Then imposing Eqs.~\eqref{k-rho}-\eqref{P2} in Eq.~\eqref{epsilon-rho}, we obtain
\be\label{sl-psi}
\beta\psi\ll 1\,, \quad \textmd{and} \quad \frac{\psi}{\beta^2}\ll 1\,.
\ee
Hereafter, we simplify and analyze the dynamical field equations assuming quasi-de Sitter inflation, in a sense that $\psi$ is given by Eq.~\eqref{psisl}, while $\dpsi/(\dot{\alpha}\psi)$, $\dot{\chi}/\dot{\alpha}$ and $\dsigma/\dot{\alpha}$ are very small during inflation (Eq.~\eqref{sl-V}).

Imposing the slow-roll conditions in Eqs.~\eqref{sl-V}-\eqref{sl-psi} and keeping the leading terms, one can simplify Eq.~\eqref{betaeq} and obtain the following form for the field equation of $\beta$ 
\bea\label{betaeq3}
\beta(\beta^6+2)(\ddbeta+3\dalpha\dbeta)+\dalpha^2\beta^2(\beta^6-1)(2+\gamma\beta^2)-6\dbeta^2\simeq0\,,
\eea
where $\gamma\equiv g^2\psi^2/\dalpha^2$.
Note that, starting with a positive (negative) $\beta$, it always remains positive (negative) during inflation\footnote{As $\beta=0$ is a singular point in Eq.~\eqref{betaeq3}, the sign of $\beta$ is fixed during slow-roll inflation.}. Moreover, using the slow-roll conditions in Eq.~\eqref{sigmaeq}, one obtains $\dot{\sigma}/\dot{\alpha}$ as
\be\label{dsigma}
\frac{\dot{\sigma}}{\dot{\alpha}}\simeq\frac{\psi^2}{3\beta^4}\bigg((1-\beta^6)-(2+\beta^6)\frac{\dot{\beta}}{\dot{\alpha\beta}}\bigg)\,.
\ee
As we already discussed, the field equations of $\beta$ and $\dot{\sigma}$ are both originated from the Yang-Mills and Einstein-Hilbert terms which are common between chromo-natural and gauge-flation models. Moreover, during the slow-roll inflation, $\psi$ and $\chi$ undergo slow-roll transitions and the Hubble parameter $\dot{\alpha}$, is almost constant. As a result, the anisotropic fields $\beta$ and $\dot{\sigma}$, have identical slow-roll behaviors in these two models. 

\begin{figure}[t]
\includegraphics[width=0.45\textwidth]{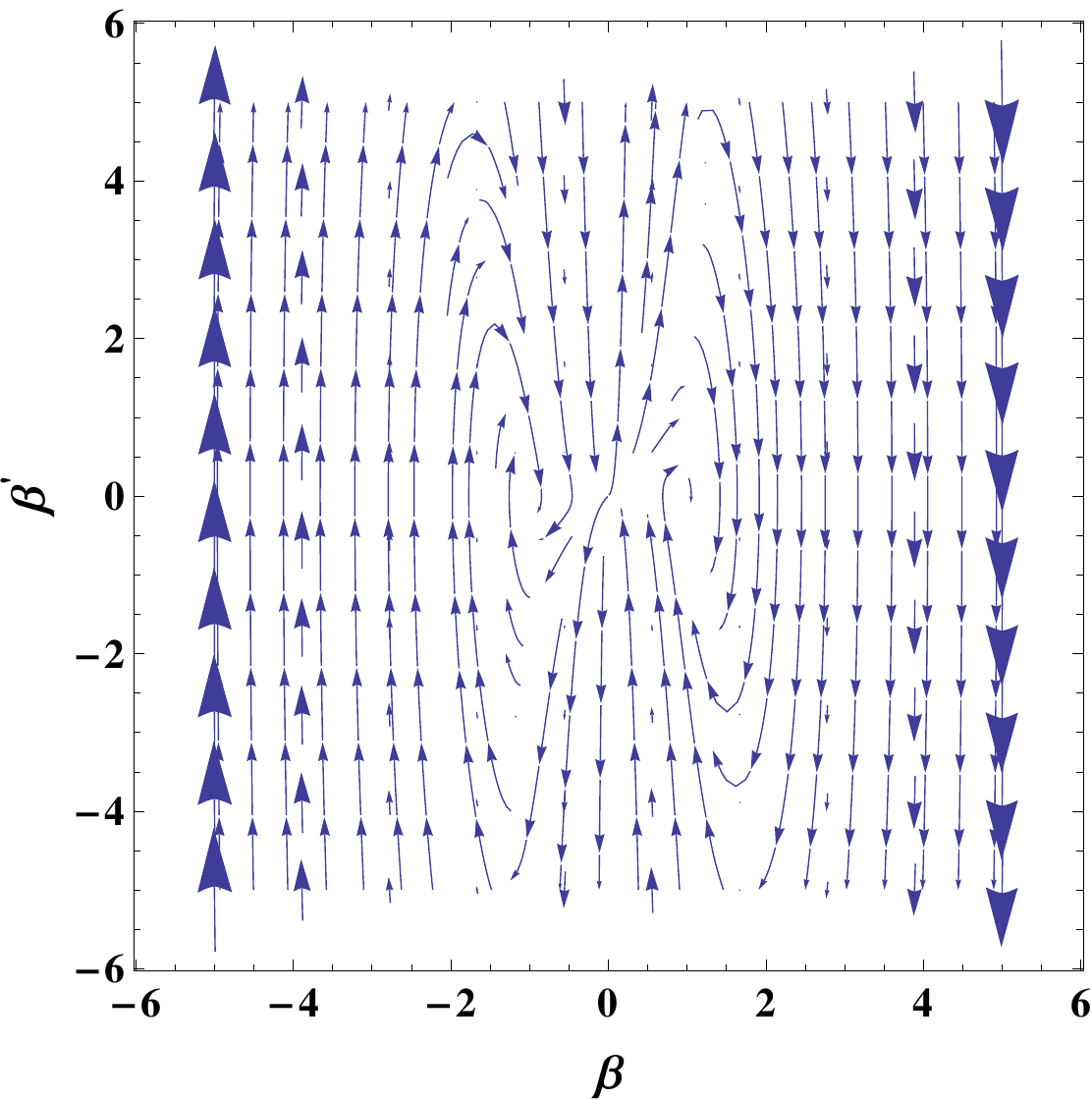}
\hspace*{.5cm}
\includegraphics[width=.47\textwidth]{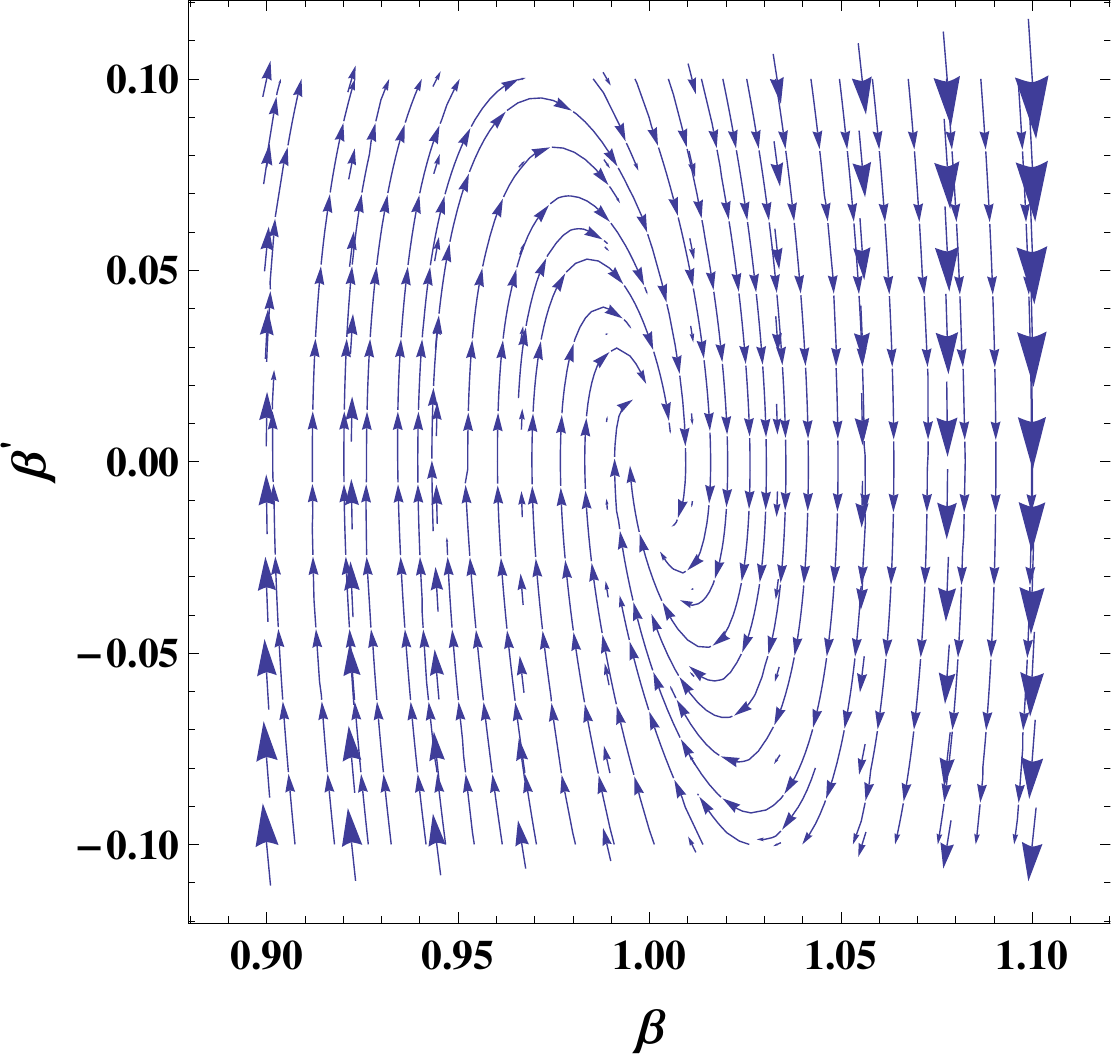}
\caption{The phase diagram in the $\beta-\beta^\prime$ plane (where $\beta'\equiv\dbeta/\dalpha$). The left figure shows the existence of two attractor solutions at $\beta =\pm 1$, corresponding to the isotropic FRW background. Moreover, it explicitly exhibits the $\beta^\prime(\beta) = -\beta^\prime(−\beta)$ symmetry. The right figure shows the phase diagram for $\beta$ in the vicinity of the attractor solution $\beta=1$. } 
\label{fig:phase}
\end{figure}

Studying Eq.~\eqref{betaeq3}, we presented the phase diagram in the $\beta-\beta^\prime$ plane (figure~\ref{fig:phase}). It is clear that  all of the trajectories approach the isotropic fixed points; \ie $\beta^2=1$. In Eq.~\eqref{betaeq3}, one can distinguish the following three regions for different values of $\beta$.

\begin{itemize}
\item  $\beta^2$ at the vicinity of one ($\beta=\pm 1+\delta\beta$, where $|\delta\beta|\ll 1$):
here, $\delta\beta$ is a damped oscillator which vanishes in one or two $e$--folds and $\beta$ exponentially approaches one of the attractor points at $\beta=\pm 1$ (isotropic configurations)
\be\label{dellamdasol}
\delta\beta\simeq 6e^{-\frac32\dot\alpha t}\bigg(A_1\cos(\sqrt{\frac74+2\gamma}~\dot\alpha t)+A_2\sin(\sqrt{\frac74+2\gamma}~\dot\alpha t)\bigg)\,.
\ee
Moreover, $\left|\dsigma\right|$ is given by
\be
\frac{\dot\sigma}{\dot{\alpha}}\simeq-\frac{\psi^2}{3}\big(\frac{1}{2}\frac{\delta\beta'}{\delta\beta}+1\big)\delta\beta\,,
\ee
which undergoes a damped oscillation, similar to $\delta\beta$.
	
\item Starting from a very small $\beta^2$, we have the following solution for $\beta$
\be\label{lambda0}
\frac{1}{\beta^2}\simeq B_1e^{-2\dot\alpha t}+B_2e^{-\dot\alpha t}\,,
\ee
which shows that $\left|\beta\right|$ is growing very rapidly and escaping quickly from the vicinity of zero. Note that this solution is only valid in first few $e$-folds where $\beta^2$ is far from one. Once $\left|\beta\right|$ gets close to one, it starts damped oscillations around the attractor solution.
As we see here, $\beta$ monotonically increases for all possible values of $B_1$ and $B_2$, but this is not necessarily the case for $\dot\sigma$. In particular, for two different initial conditions in which (i) $B_1=0$ and (ii) $B_2=0$, we have
\be\label{sigma-s}
\frac{\dot\sigma}{\dot{\alpha}}|_{B_1=0}\simeq-\frac{\beta_0^2\psi^2}{2}e^{\dot\alpha t}\,, \quad \textmd{and}\quad \frac{\dot\sigma}{\dot{\alpha}}|_{B_2=0}\simeq-\frac{\psi^2}{3\beta_0^4}e^{-4\dot\alpha t},
\ee
where $\beta_0$ is the initial value of $\beta$. 
\item Initiating from a very large $\left|\beta\right|$, we have
\be\label{beta-l}
\beta\simeq C_1e^{-2\dot\alpha t}+C_2e^{-\dot\alpha t}\,,
\ee
which has an exponentially damping behavior and when $\left|\beta\right|$ gets close to one, it starts damped oscillations around the attractor solution. Note that Eq.~\eqref{beta-l} is identical to Eq.~\eqref{lambda0} that governs the evolution of $1/\beta^2$ in the limit $\left|\beta\right|\ll 1$. For all possible values of $C_1$ and $C_2$ in \eqref{beta-l}, $\beta$ monotonically decreases, however this is not necessarily the case for $\dot\sigma$. In fact, for two initial conditions in which (i) $C_1=0$ and (ii) $C_2=0$, we have
\be\label{sigma-l}
\frac{\dot\sigma}{\dot{\alpha}}|_{C_1=0}\simeq\frac{\psi^2}{\beta_0^4}e^{4\dot\alpha t}\,, \quad \textmd{and}\quad \frac{\dot\sigma}{\dot{\alpha}}|_{C_2=0}\simeq\frac{\psi^2\beta_0^2}{3}e^{-4\dot\alpha t}\,,
\ee
where $\beta_0$ is the initial value of $\beta$.
\end{itemize}
	
\begin{figure}[h!]
\centering
\includegraphics[width=1\textwidth]{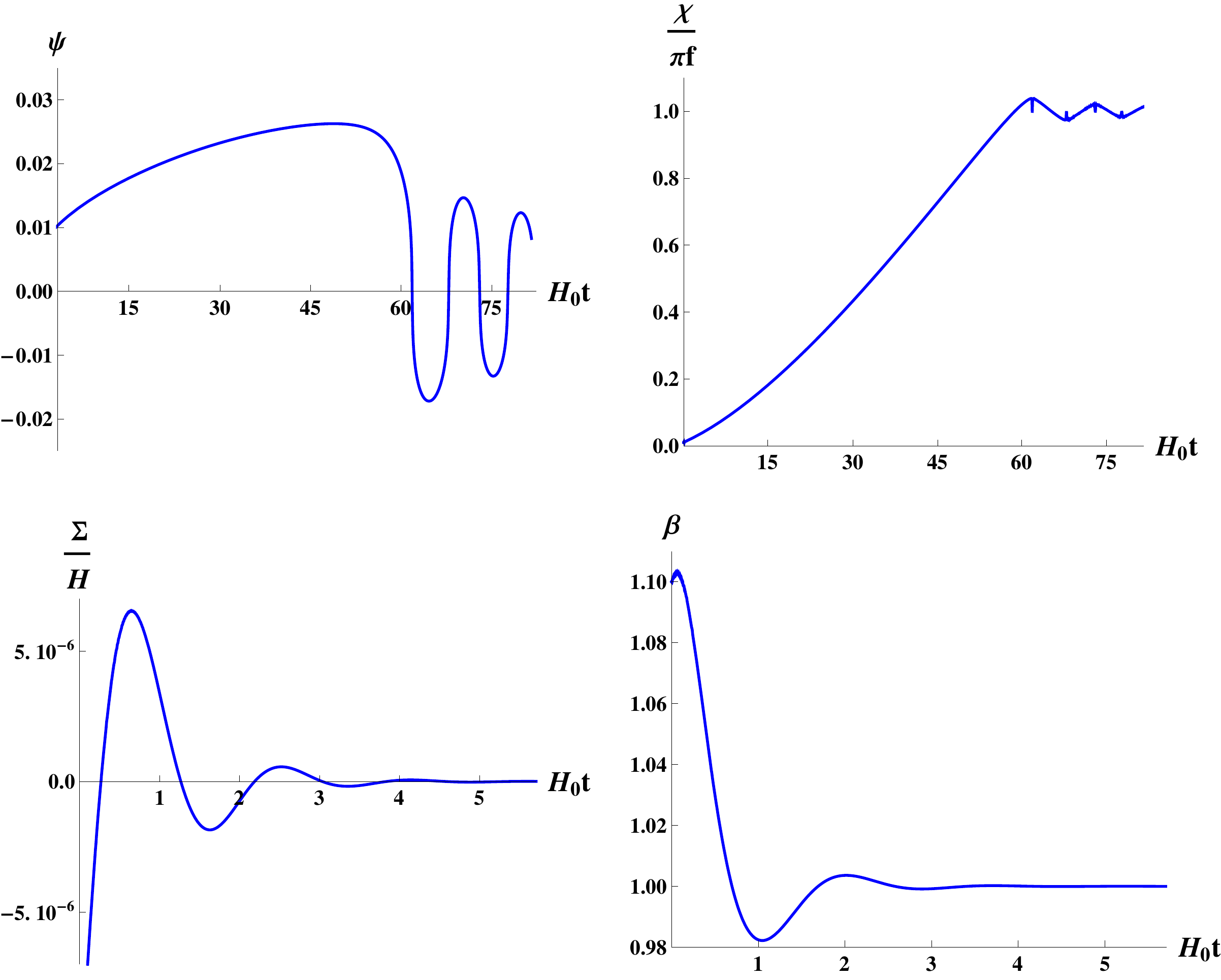}
\caption{The classical trajectories of $\psi$ (top left), $\chi/(\pi\,f)$ (top right), $\dsigma/\dalpha\equiv\Sigma/H$ (bottom left) and $\beta$ (bottom right) for the parameters $\lambda=200,\,\mu=10^{-4},\,f=0.01\,,g=2\times10^{-6}$. The initial values are $\beta_0=1.1$, $\psi_0=10^{-2}$ and $\chi_0/(f\pi)=10^{-2}$.}
\label{fig:all}
\end{figure}
	
In a system which undergoes quasi-de Sitter inflation (\ie very small $\epsilon$), we see that regardless of initial $\beta$ values, all solutions converge to $\beta^2 = 1$ (the isotropic solutions), within the first few $e$--folds. In other words, the isotropic inflation is the attractor solution in the chromo-natural model. Furthermore, we demonstrate that initiating from $|\beta|\gg1$ ($|\beta|\ll1$), $\beta$ decreases (increases) exponentially fast and gets close to one in few $e$--folds.
Interestingly, this is not necessarily the case for $\dot\sigma$. From Eqs.~\eqref{sigma-s} and \eqref{sigma-l}, we realize that having a gauge (vector) field in the system, $|\dot{\sigma}/\dot{\alpha}|$ can grow for one or two $e$--folds during the slow-roll inflation and saturates its theoretical upper bound $|\dot{\sigma}/\dot{\alpha}|\sim\epsilon$ \cite{Maleknejad:2012as}\footnote{Note that in the case of scalar field models in context of GR, this is not possible. In scalar models, the anisotropies of the metric are source-free and always damp  exponentially fast during the slow-roll inflation \cite{Maleknejad:2012as}.}. However, $\beta$ gets close to one quickly, and eventually $|\dot{\sigma}/\dot{\alpha}|$ starts damping exponentially fast after the first few $e$--folds.
\vskip 0.5 cm
\textit{Numerical analysis:}\\
	
At this point, we present the numerical analysis of the system. We have studied the system for different values of $\beta$, and different $\chi/f$ values numerically. Note that, demanding slow-roll inflation, one can choose the initial value of $\chi$ as $\chi_0/f\in(0,\pi)$, where $\chi_0$ is the initial value of the axion field \cite{Maleknejad:2012dt}. In the limit that $\chi_0/f\simeq\pi$, the chromo-natural model reduces to the gauge-flation model \cite{SheikhJabbari:2012qf}. 
Our numerical study also confirms the analytical results. Here, we present the classical trajectories for one set of values in figure~\ref{fig:all}. The parameters are $\lambda=200,\,\mu=10^{-4},\,f=0.01\,,g=2\times10^{-6}$, while initial values are $\beta_0=1.1$, $\psi_0=10^{-2}$ and $\chi_0/(f\pi)=10^{-2}$.
The top left figure shows the classical trajectory of the field $\psi$ with respect to $H_0 t$, while the top right figure indicates dynamics of the axion field, $\chi$. It is clear that there is a period of quasi-de Sitter inflation, where $\psi$ and $\chi$ remains almost constant. In fact, the anisotropic slow-roll behaviors of $\psi$ and $\chi$ are essentially the same as their corresponding fields in the isotropic setup \cite{Maleknejad:2012dt}. The bottom left and right figures show evolutions of two dimensionless variables, $\dsigma/H$ and $\beta$ during the first several $e$--folds, respectively. The left bottom figure indicates that $\Sigma/H$ decreases quickly and becomes negligible within few $e$--folds and then the system mimics the behavior of isotropic inflation. The right bottom figure shows $\beta$, initially started from $\beta_0 = 1.1$, quickly decreases and gets close to the isotropic solution, $\beta=1$.

\section{Conclusions}\label{V}

Considering the chromo-natural model in the Bianchi type I background, we have studied the classical stability of isotropic slow-roll inflation with respect to initial anisotropies. For the inflationary models which consist of (only) scalar fields, the anisotropies are (exponentially) damped due to the inflationary expansion of the universe and they respect the cosmic no-hair conjecture \cite{Wald:1983ky}. However, this is not necessary the case in the presence of vector (gauge) fields. More precisely, \textit{inflationary extended cosmic no-hair theorem} \cite{Maleknejad:2012as} states that for general inflationary systems of all Bianchi types, in contrast to the cosmic no-hair conjecture (and also Wald's cosmic no-hair theorem\cite{Wald:1983ky}), anisotropies can in principle grow during inflation. However, assuming a (slow-roll) quasi-de Sitter expansion, inflation puts an upper bound on the enlargement of the anisotropies and enforces them to be of the order of the slow-roll parameter $\epsilon$.

We parametrized the anisotropy in the metric with $\sigma(t)$, while $\beta(t)$ is a parameter that represents anisotropy in the gauge field and the isotropic setup is described by $\dot\sigma=0$ and $\beta=\pm1$. We showed that although the gauge field is turned on in the background, the isotropic solution is an attractor solution in the chromo-natural model and the anisotropies are damping within first few $e$--folds. Here, for simplicity in the analytical study, we considered an axial symmetric version of Bianchi type I. However, our numerical investigations confirmed that this attractor behavior is a generic property of the system, \textit{i.e.} the attractor domain includes all the initial anisotropies.
In other words, the chromo-natural model respects the cosmic no-hair conjecture. 
In particular, we demonstrated that starting from $|\beta|\gg1$ ($|\beta|\ll1$), $\beta$ decreases (increases) exponentially fast and gets close to the isotropic configuration in few $e$--folds.
However, this is not necessarily the case for $\dot\sigma$. Since we have a gauge (vector) field in the system, $|\dot{\sigma}/\dot{\alpha}|$ may increase for one or two $e$--folds during the slow-roll inflation and saturates its theoretical upper bound $|\dot{\sigma}/\dot{\alpha}|\sim\epsilon$ \cite{Maleknejad:2012as}. However, as $\beta$ approaches one, $|\dot{\sigma}/\dot{\alpha}|$ damps exponentially fast after the first few $e$--folds.

We showed that the anisotropies in both chromo-natural and gauge-flation models share the same dynamics during the slow-roll inflation. In \cite{Maleknejad:2012fw}, it has been shown that these two models share the same vector (and also tensor) mode perturbations around the isotropic and homogeneous FRW background. The reason for the above correspondence between these two models is that, both vector perturbations and the background anisotropies are originated from the Yang-Mills term in the action which is common in chromo-natural and gauge-flation models. Moreover, during the slow-roll inflation, $\psi$ and $\chi$ undergo slow-roll transition and the Hubble parameter $\dot{\alpha}$, is almost constant. In fact, in both chromo-natural and gauge-flation models, 
the quasi-de Sitter expansion is globally stable against the initial anisotropies and the anisotropies are damped within first few $e$--folds, in agreement with the cosmic no-hair conjecture. 
 
\section*{Acknowledgments}
We would like to thank M. M. Sheikh-Jabbari for very useful comments on the draft and Angelo Ricciardone for his collaboration in the early stage of this work.
AM and EE acknowledge the Abdus Salam ICTP for hospitality where part of this work has been done. 
The final stage of this work was carried out during AM visit to the Max Planck Institute for Astrophysics and she greatly appreciates Eiichiro Komatsu for his warm hospitality and fruitful discussions.
EE also acknowledges partial support from the European Union FP7 ITN INVISIBLES (Marie Curie Actions, PITN-GA-2011-289442). AM is supported in part by the grants from Boniad Melli Nokhbegan of Iran.

\end{document}